\documentclass[10pt,twocolumn,letterpaper]{article}

\usepackage{wacv}              

\usepackage{graphicx}
\usepackage{amsmath}
\usepackage{amssymb}
\usepackage{booktabs}
\usepackage{etoolbox}
\usepackage{multirow}
\usepackage{caption}
\usepackage{enumitem}
\usepackage{dblfloatfix}
\usepackage{pifont}
\usepackage{bm}
\newcommand{\xmark}{\ding{55}}%

\usepackage[pagebackref,breaklinks,colorlinks]{hyperref}

\usepackage[capitalize]{cleveref}
\crefname{section}{Sec.}{Secs.}
\Crefname{section}{Section}{Sections}
\Crefname{table}{Table}{Tables}
\crefname{table}{Tab.}{Tabs.}

\def\wacvPaperID{1053} 
\def\confName{WACV}
\def\confYear{2025}

\begin{document}

\setlength{\abovedisplayskip}{4pt}
\setlength{\belowdisplayskip}{4pt}
\title{SyncViolinist: Music-Oriented Violin Motion Generation\\
Based on Bowing and Fingering}

\author{Hiroki Nishizawa\\
Waseda University\\
Tokyo, Japan\\
{\tt\small pipidesu2430@akane.waseda.jp}
\and
Keitaro Tanaka\\
Waseda University\\
Tokyo, Japan\\
{\tt\small phys.keitaro1227@ruri.waseda.jp}
\and
Asuka Hirata\\
Waseda University\\
Tokyo, Japan\\
{\tt\small asuka112358@suou.waseda.jp}
\and
Shugo Yamaguchi\\
Waseda University\\
Tokyo, Japan\\
{\tt\small wasedayshugo@suou.waseda.jp}
\and
Qi Feng\\
Waseda Research Institute for Science and Engineering\\
Tokyo, Japan\\
{\tt\small feng@aoni.waseda.jp}
\and
Masatoshi Hamanaka\\
RIKEN\\
Tokyo, Japan\\
{\tt\small masatoshi.hamanaka@riken.jp}
\and
Shigeo Morishima\\
Waseda Research Institute for Science and Engineering\\
Tokyo, Japan\\
{\tt\small shigeo@waseda.jp}
\\ \bigskip \insertfig
}
\maketitle

\begin{abstract}
Automatically generating realistic musical performance motion can greatly enhance digital media production, 
 often involving collaboration between professionals and musicians. However, capturing the intricate body, hand, and finger movements required for accurate musical performances is challenging. Existing methods often fall short due to the complex mapping between audio and motion, typically requiring additional inputs like scores or MIDI data.
In this work, we present SyncViolinist, a multi-stage end-to-end framework that generates synchronized violin performance motion solely from audio input. Our method overcomes the challenge of capturing both global and fine-grained performance features through two key modules: a bowing/fingering module and a motion generation module. The bowing/fingering module extracts detailed playing information from the audio, which the motion generation module uses to create precise, coordinated body motions reflecting the temporal granularity and nature of the violin performance.
We demonstrate the effectiveness of SyncViolinist with significantly improved qualitative and quantitative results from unseen violin performance audio, outperforming state-of-the-art methods. Extensive subjective evaluations involving professional violinists further validate our approach.
The code and dataset are available at 
\url{https://github.com/Kakanat/SyncViolinist}.
\vspace{-4mm}
\end{abstract}


\begin{figure}[hbtp]
\includegraphics[width=0.99\linewidth]{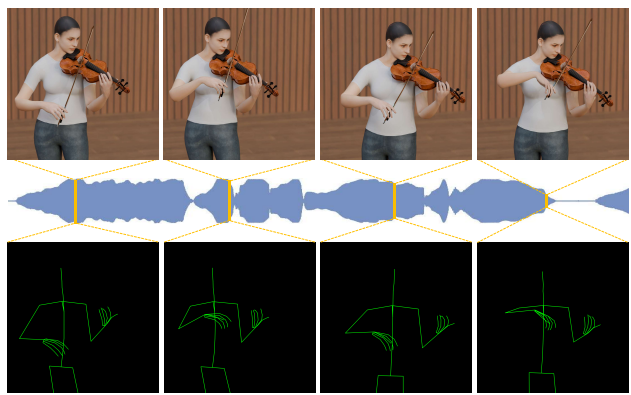} 
  \caption{SyncViolinist can automatically generate synchronized violin performance motions entirely from the audio input, accurately reflecting global and fine-grained performance features such as natural body movements and coordinated bowing and fingering.}
  \vspace{-5mm}
\label{fig:teaser}
\end{figure}

\section{Introduction}
Generating realistic and natural character motions is crucial to modern digital media production and virtual reality experiences. To achieve high realism, traditional methods are often costly and labor-intensive, involving specialized techniques such as hand-crafting animations and time-consuming motion capture processes. This is particularly true for musical performances, where the consultation of specialized artists is often necessary during the hand-crafting or post-processing, disrupting the production flow.

A framework that allows the automatic generation of natural and realistic motion in musical performance from only audio has the potential to improve workflow efficiency significantly and is highly desirable. 
This paper focuses on violin performance and aims to generate body motions from audio signals.
In violin performance, 
the primary hand motions that affect the sound produced on the instrument are 
\textit{bowing} and \textit{fingering}. 
Bowing, which is executed by the right hand, involves the movement of the bow across the strings, whereas fingering, which is performed by the left hand, involves pressing the strings on the frets to control pitch. 
The performer's interpretation of the musical piece is reflected in the sounds and coordinated body motions through bowing and fingering.

Creating an automatic system, however, is a challenging task as it requires capturing both natural global body motions and precise fine-grained hand and finger motions. Previous rule-based~\cite{Yin:05} and template-based~\cite{Serafin:01, Zhu:13, Yamamoto:10} approaches often require additional inputs, such as corresponding musical scores or MIDI information, which limits their practicality. While learning-based methods have attempted to establish the correspondence between motion and audio in an end-to-end manner~\cite{Shlizerman:18, Kao:20, Chen:21}, they often fail to capture the fine features present in musical performance. This is primarily due to the complex and unconstrained correspondence between sound and body motions, leading to sub-optimal motion generation.

In this paper, we propose \textit{SyncViolinist}, a multi-stage end-to-end framework for generating synchronized violin performance motion solely from audio input (Fig.~\ref{fig:teaser}). 
Our approach overcomes the challenge of establishing global and fine correspondences for natural motion generation through a two-stage process: a \textit{bowing/fingering module} and a \textit{motion generation module}.
We employ convolutional recurrent neural networks (CRNNs) to predict the bowing and fingering from the Mel-scaled spectrogram extracted from the input audio.
Subsequently, our novel motion generation module utilizes multiple parallel bidirectional long short-term memory (BiLSTM) branches. This stage considers both audio features and fine-grained playing information provided by the CRNNs to predict nuanced global-fine motion patterns for various body semantics.
Due to the bowing/fingering passed to the motion generation module, we narrow down the infinite patterns of global body motions to constrained combinations.
To facilitate integrated learning,
we present a new dataset obtained from professional musicians, consisting of synchronized audio, body motions, and accurately annotated bowing/fingering information.

The experimental results of this study demonstrate that the proposed approach can accurately estimate a time series of joint positions from audio signals with significantly improved performance when compared to state-of-the-art methods. An ablation study further demonstrates the effectiveness of incorporating bowing/fingering information and body semantics. 
Finally, we conducted an extensive subjective evaluation with over 40 participants, including four professional violinists, 
revealing that the generated motion possessed a higher level of realism.

The main contribution of this work is two-fold;
\setlist{nolistsep}
\begin{itemize}[noitemsep]
    \item We propose a novel multi-stage, end-to-end framework that generates realistic and natural motion for violin performance solely from audio signals via bowing/fingering information;
    \item We construct an entirely new dataset of violin performances by professional musicians with synchronized audio signals, body motions, and accurately annotated bowing/fingering information, which will be released with the codes for future research upon acceptance. 
\end{itemize}

\begin{figure*}[hbtp]
\includegraphics[width=0.99\linewidth]{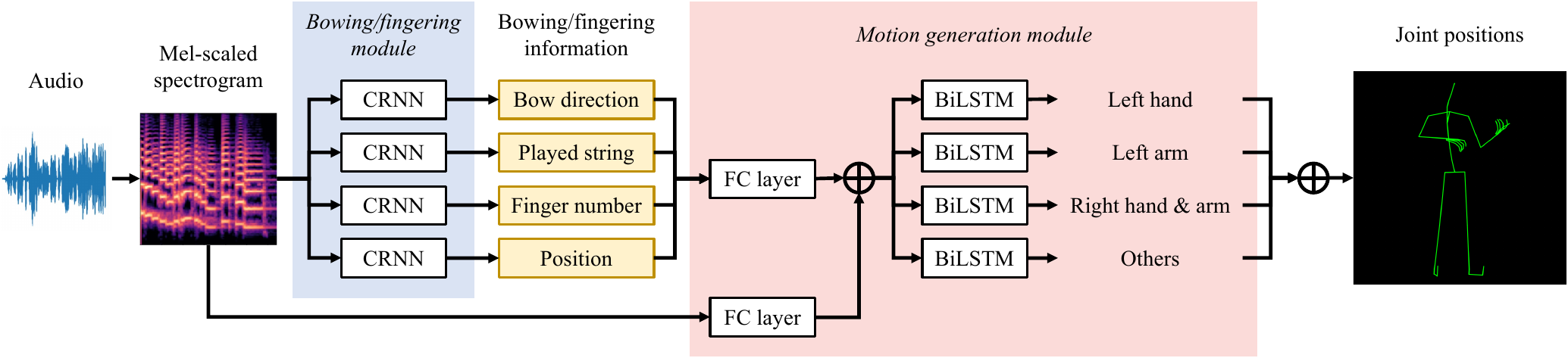} 
  \vspace{-1mm}
  \caption{Proposed method overview. The framework has two components: a bowing/fingering module and a motion generation module.}
  \vspace{-5mm}
\label{fig:overview}
\end{figure*}

\section{Related Work}

In this section, we first overview recent research related to the motion generation task using audio signals as input. This is followed by a more in-depth examination of the history of motion synthesis, specifically for musical instrument performance. Finally, we review the existing efforts to analyze the actions of musicians while playing instruments.
    
\subsection{Audio-based motion generation}

Audio-based motion generation is a widely researched area in fields such as facial expressions, conversational gestures, dance, conducting, and musical performance. In recent years, deep learning techniques have been widely employed in these fields. 

For facial animation, researchers have used CNN to examine the correlation between audio and facial expressions~\cite{Karras:17}. LSTM can also capture temporal dependencies~\cite{Pham:17} and enhance the performance of the model~\cite{Pham:18}.
 Similarly, LSTM-based structures have been used for conversational gestures 
  to generate 3D joint angles and 2D joint positions from speech~\cite{Ferstl:18, Ginosar:19}. 
 Kucherenko et al.~\cite{Kucherenko:19} employed an encoder-decoder structure to map 3D joint position into a pose embedding space and then utilized an LSTM-based framework to regress the pose embedding.

For music-related animation, the task of dance motion generation has gained significant attention. Various methods have been proposed, including the use of factored conditional restricted Boltzmann machines~\cite{Alemi:17}, transformers~\cite{Li:21ai}, and dilated convolution-based models~\cite{Zhuang:22}. Recently, contrastive learning have been used to optimize music encoders in generative models, resulting in synchronized motion with input music~\cite{Liu:22}. 
Diffusion-based models have also succeeded in audio-driven generation with their strong ability to learn complex dynamics~\cite{zhao2023taming, zhu2023taming}. 
However, they require large-scale quality performance as input, 
which limits their viability in applications due to data inefficiency.


Furthermore, it is important to note that even though such motion generation tasks can be applied to musical performance, naive approaches are often unable to provide satisfying results.
This is because musical motions are often dictated by scores and also constrained by techniques~\cite{zhao2023taming} and interaction with instruments~\cite{young2007bowstroke}.

\subsection{Motion synthesis for musical performance}

This section explores the various efforts to generate body motions for musical instrument performance. Different approaches range from cost-minimization \cite{Elkoura:03} and rule-based~\cite{Yin:05} to template-based solutions~\cite{Serafin:01, Zhu:13, Yamamoto:10}. Recently, researchers have employed deep learning methods to generate motion in an end-to-end manner. Li et al.~\cite{Li:18} used MIDI as input for a combined CNN and RNN model to generate the body motions of a pianist. While MIDI information is helpful for keyboard instruments, it is less applicable for violin as MIDI-equipped violins are not as common. 

An alternative approach is to adopt audio-driven methods. Shlizerman et al.~\cite{Shlizerman:18} generated body motions of piano and violin performance from audio by estimating 2D joint positions of each body part using LSTM. Kao and Su~\cite{Kao:20} then designed a two-branch network to generate the motions of the right hand and other body parts, taking into account the characteristics of fine-grained motions required for violin bowing. 
Chen et al.~\cite{Chen:21} proposed a generative adversarial network (GAN) to generate upper body motions based on the performance audio of Guzheng.
While previous research has succeeded in generating motions in an end-to-end manner, they did not consider how musicians plan their motions, such as fingering, which is an essential aspect of creating intended sounds during performance. 

\subsection{Analysis of musical performance motions}

In musical performance, the body motions of performers play an important role in reflecting the sound~\cite{Dalmazzo:19} and enhancing its expressiveness and engagement with the audience~\cite{Broughton:09}. 
Research has shown that they varies among performers to embody the intended sound and musical structure~\cite{Davidson:12, Macritchie:13}. For violin performance, the choice of bowing and fingering techniques results in different body motions. As a result, the relationship among bowing, fingering, and violinists' body motions has been widely studied. 
Dalmozzo and Ram{'\i}rez~\cite{Dalmazzo:19} classified bowing techniques 
based on hierarchical hidden Markov models and motions from a professional violinist. 
Baader et al.~\cite{Baader:05} studied the synchronization of the right- and left-hand motions, which are respectively responsible for bowing and fingering.

There have been several works that focus on 
string instruments, such as the violin, specifically on the topics of fingering and bowing. 
To determine fingering, one of the most crucial issues for artists, previous methods~\cite{Jen:21,Cheung:21semi} estimate the strings, finger numbers, and positions to be used from scores. Maezawa et al.~\cite{Maezawa:10} proposed to estimate the strings and finger numbers from the score and acoustic information using the differences in timbre depending on the strings used. 
In contrast, our method uses bowing and fingering information obtained entirely from audio signals.

\section{Method}
\label{sec:3}

In violin performance, artists interpret the music through their body motions, which are closely tied to the chosen bowing and fingering techniques that produce specific sounds. Therefore, incorporating explicit bowing and fingering information into motion generation enhances realism and efficiency compared to relying solely on audio inference. This approach narrows the range of possible motions, aligning closely with the performer’s physical execution. Moreover, different body parts (such as the left hand and the right hand) exhibit varying temporal granularity during playing. Thus, separating body semantics when training the motion generation module facilitates the creation of natural, coordinated global and fine-grained motions. With these considerations in mind, we propose the following approach.

    
An overview of the proposed method is shown in Fig.~\ref{fig:overview}.
Our proposed method consists of two components:
    a \textit{bowing/fingering module} and a \textit{motion generation module}.
The bowing/fingering module estimates four types of features: bow direction, played string, finger number, and position, which dictate violinists' bowing and fingering. These four features are defined as bowing/fingering information.
The motion generation module uses the bowing/fingering information and audio features as inputs to estimate the time series of joint positions for each body semantic in groups.


\subsection{Bowing/fingering module}
\newcommand{\argmax}{\mathop{\rm arg~max}\limits}
\label{subseq:3_1}
The bowing/fingering module consists of four networks,
    all of which take Mel-scaled spectrogram 
    $X^{\mathrm{mel}}=x^{\mathrm{mel}}_{1:T, 1:F}\in \mathbb{R}^{T \times F}$
    as input,
    where $T$ is the number of total time frames,
    and $F$ is the number of frequency bins.
Each of them predicts the time series of each of the four features
    in a one-hot representation,
    bow direction $L^{\mathrm{bow}}=l^{\mathrm{bow}}_{1:T,1:3}\in \{0,1\}^{T \times 3}$,
    played string $L^{\mathrm{str}}=l^{\mathrm{str}}_{1:T,1:5}\in \{0,1\}^{T \times 5}$,
    finger number $L^{\mathrm{fing}}=l^{\mathrm{fing}}_{1:T,1:6}\in \{0,1\}^{T \times 6}$,
    and position $L^{\mathrm{pos}}=l^{\mathrm{pos}}_{1:T,1:13}\in \{0,1\}^{T \times 13}$,
    where the second dimension indicates the class number.
Here,
    bow direction class $\{1,2\}$ indicates \{up, down\}-bow,
    the main two bow mode.
Played string $\{1,2,3,4\}$ indicates each of the four strings of the violin,
    \{E, A, D, G\} string, 
    from the highest pitch to the lowest.
Finger number $\{1,2,3,4,5\}$ indicates 
    \{open string (\textit{i.e.}, no pressed down finger), index, middle, ring, pinky\} finger.
Position $\{1,2,3, ...,12\}$ indicates the \{first, second, third, ..., twelveth\} position.
In violin playing, ``position'' refers to the placement of the left hand on the fingerboard of the violin to produce different notes. Each position corresponds to a specific range of notes that can be played on the instrument. Additionally, the last class in each feature indicates silence periods.
    

All four components in our proposed method are built upon CRNNs, which have been demonstrated to be effective for sound event detection tasks ~\cite{Adavanne:17}. We propose an architecture that combines three blocks of a 2D convolutional network, two layers of BiLSTM ~\cite{Graves:05}, and two fully connected (FC) layers in each of our four branches.
Each convolutional layer is followed by 2D batch normalization, a 2D max-pooling layer, and Leaky ReLU activation functions~\cite{Maas:13}. The BiLSTM layers are followed by a dropout layer to regularize the network. The subsequent FC layer utilizes a Leaky ReLU activation function, and the final FC layer is activated by a softmax function. Each of the four networks outputs the probability of the corresponding bowing/fingering feature belonging to a specific class at each time frame
    $P^{f}= p^{f}_{1:T,1:n_{f}} \in (0,1)^{T \times n_{f}}$,
    where $f$ indicates the feature type $\{\mathrm{bow}, \mathrm{str}, \mathrm{fing}, \mathrm{pos}\}$,
    and $n_f$ indicates the total number of classes for each feature
    (\textit{i.e.}, $\{n_{\mathrm{bow}}, n_{\mathrm{str}}, n_{\mathrm{fing}}, n_{\mathrm{pos}}\}=\{3,5,6,13\}$).

The proposed method utilizes a multi-branch architecture, each component focusing on different audio features to extract the corresponding bowing/fingering feature. 
The final output is obtained in a one-hot format by selecting the class with the highest output probability at each time frame:
\begin{equation}
    l^{f}_{t,c} = 
    \begin{cases}
    1 & \mathrm{if} ~p^{f}_{t,c} = \underset{c'}{\max}\;{p^{f}_{t,c'}}\\
    0 & \mathrm{otherwise.}
    \end{cases}
\end{equation}
In the training process, 
    we minimize the cross-entropy loss $\mathcal{L}_{\mathrm{ce}}$ for each output
    represented as follows:
\begin{equation}
\mathcal{L}_{\mathrm{ce}} = - \frac{1}{T}\sum_{t=1}^{T}{\sum_{c=1}^{n_{f}}{\hat{p}^{f}_{t,c}}\cdot \log{p^f_{t,c}}},
\end{equation}
where $\hat{p}^{f}_{t,c}$ is the ground truth probability for the bowing/fingering feature
    class $c\in\{1,...,n_f\}$ at time frame $t$.

\begin{table*}[t]
    \centering
    \setlength\tabcolsep{1.3mm}
    \resizebox{1.8\columnwidth}{!}{%
    \begin{tabular}{lccccccccc}
    \toprule
     Dataset & Instrument & Duration & Pieces & Joints & Marker/Sensor & Modality \\
     \midrule
    Young and Deshmane~\cite{young2007bowstroke} & Violin & N/A & N/A & N/A & \checkmark & Bow\\
    Marchini et al.~\cite{marchini2014sense} & String quartet & N/A & 23 & N/A & \checkmark & Bow\\
    Volpe et al.~\cite{volpe2017multimodal} & Violin & 2.4 h & 41 & 48  & \checkmark & Body, Instrument, Bow\\ 
    Jin et al.~\cite{jin2024audio} & Violin and Cello & 3.0 h & 120 & 140 & \xmark & Hands, Body, Instrument, Bow\\
    Shlizerman et al.~\cite{Shlizerman:18} & Violin & 3.6 h & N/A & 25 & \xmark & Hands, Body\\
    Kao and Su~\cite{Kao:20} & Violin & 11 h & 14 & 15 & \xmark & Body\\
    {\bf {Ours}} & Violin & 3.1 h & 61 & 75 & \checkmark & Hands, Body, Bow\\
\bottomrule
    \end{tabular}
    }
        \vspace{-2mm}
        \caption{Comparison of commonly used violin performance datasets. The proposed dataset has the most pieces and tracked joints among captured datasets acquired with markers and sensors.}
        \label{table:dataset_comparison}
    \vspace{-6mm}
\end{table*}

\subsection{Motion generation module}
The motion generation module is designed to generate a time series of body joint positions 
 from audio signals and the estimated bowing/fingering information. 
First, the bowing/fingering information $L^{\mathrm{bf}}\in \{0,1\}^{T \times 27}$,
 obtained by concatenating all $L^{f}$ estimated in Section~\ref{subseq:3_1}, and the audio features $X^{\mathrm{mel}}$ are separately passed through FC layers to obtain embedded features.
The outputs of the FC layers are activated by a Leaky ReLU function.

The embedded bowing/fingering information and audio features are then concatenated
 and fed into multiple parallel BiLSTM branches, each specializing in different body parts. 
This separation, based on the varying temporal granularity during violin performance and body semantics, results in distinct branches for the left hand, left arm, right hand \& arm, and other body parts.
Each BiLSTM branch is followed by a dropout layer and an FC layer,
 producing a time series of 3D joint positions for the corresponding body parts.
The four groups of 3D joint positions are then concatenated,
 resulting in a time series of 3D body joint positions $J={\bm {j}}_{1:T,1:N} \in \mathbb{R}^{T \times N \times 3}$, where $N$ is the total number of joints (set to 75 in our dataset).
This multi-branch design, combined with the bowing/fingering information and audio features,
 allows us to achieve both natural global body motions hinted by input audio features 
 and precise fine-grained hand and finger motions guided by explicit bowing/fingering information.

In the training process,
    the module is supervised by two functions: joint position loss and displacement loss.
The joint position loss $\mathcal{L}_{\mathrm{jp}}$ is the L1-norm distance 
    between the generated joint positions $J$ 
    and those of the ground truth $\hat{J}={\hat{\bm {j}}}_{1:T,1:N} \in \mathbb{R}^{T \times N \times 3}$, 
    represented as follows:
\begin{equation}
\label{eq: jp}
\mathcal{L}_{\mathrm{jp}} = \frac{1}{T}\sum_{t=1}^{T}\sum_{n=1}^{N}{||{\bm {j}}_{t,n} - \hat{\bm {j}}_{t,n}||_{1}}.
\end{equation}
$\mathcal{L}_{\mathrm{jp}}$ in our experiments contributes to governing the accuracy of joint positions
 similar to the existing works~\cite{Shlizerman:18, Kao:20}.
However, since $\mathcal{L}_{\mathrm{jp}}$ alone would cause jittering in the output,
 we also include a displacement loss $\mathcal{L}_{\mathrm{dis}}$ 
 in order to guarantee temporal consistency as in ~\cite{Tevet:22}, computed as
\begin{equation}
\label{eq: dis}
\mathcal{L}_{\mathrm{dis}} = \frac{1}{T}\sum_{t=1}^{T-1}{\sum_{n=1}^{N}
 {||({\bm {j}}_{t+1,n}-{\bm {j}}_{t,n})-(\hat{{\bm {j}}}_{t+1,n}-\hat{{\bm {j}}}_{t,n})||_{1}}}.
\end{equation}
The total training objective to minimize is given as
\begin{equation}
\label{eq: ce}
\mathcal{L} = \mathcal{L}_{\mathrm{jp}} + \lambda \mathcal{L}_{\mathrm{dis}},
\end{equation}
 where $\lambda$ represents the weight assigned to the displacement loss in relation to the joint position loss.

\subsection{Post-processing}


For visualization purposes, we dress up the skeleton by optimizing the SMPL-X \cite{SMPL-X:2019} parameters, specifically the body shape parameter $\beta$ and the pose parameter $\theta$ for the body and hands/arms. 
The optimization process minimizes $E(\theta; \beta, J)$ between the estimated joint positions $J$ and the SMPL-X model's joint positions with respect to $\theta$.
$E(\theta; \beta, J)$ is calculated as
\begin{equation}
    E(\theta; \beta, J) = || R_\theta(J_{\mathrm{rp}}(\beta)) - J ||_2^2,
\end{equation}
 where $J_{\mathrm{rp}}(\cdot)$ gives the joint positions of the rest pose, 
 and $R_{\theta}(\cdot)$ transforms the given joint positions based on $\theta$. 
We use PyTorch and L-BFGS \cite{nocedal2006nonlinear} with the strong Wolfe line search.
To optimize the body pose, we employ VPoser\cite{SMPL-X:2019}, a variational autoencoder that has learned a prior distribution of poses.
We optimize VPoser's latent variables and use its output as the SMPL-X body pose parameters $\theta$.

To attach the violin and bow to the avatar,
 we use Blender constraints to make the instruments follow the SMPL-X model's joint positions.
The violin body follows the neck position and orients towards the first joint of the left thumb.
The bow follows the position of the right ring finger and points towards a point on the violin strings.

\begin{table*}[t]
    \centering
    \resizebox{1.8\columnwidth}{!}{%
    \begin{tabular}{lccccccccc}
    \toprule
     Method & L1 & L1RA & L1LA & L1LF & DTW & DTWRA & DTWLA & DTWLF & Jerk\\
     \midrule
      Shlizerman et al.~\cite{Shlizerman:18} & 21.11 & 0.60 & 0.56 & 1.40 & 30.08 & 0.94 & 0.75 & 1.84 & 550.73\\ 
Kao and Su~\cite{Kao:20} & 14.94 & 0.44 & 0.35 & 1.01 & 17.82 & 0.55 & 0.36 & 1.21 & 19399.31\\ 
Chen et al.~\cite{Chen:21} & 13.50 & 0.41 & 0.23 & 0.78 & 13.37 & 0.29 & \bf{0.13} & \bf{0.56} & 307.35\\ 
\bf{Ours} & \bf{9.09} & \bf{0.26} & \bf{0.23} & \bf{0.65} & \bf{7.62} & \bf{0.22} & 0.17 & 0.58 & \bf{298.50}\\
\bottomrule
    \end{tabular}}
    \vspace{-1mm}
        \caption{Quantitative results of comparison with baseline methods. A lower score indicates a better result.}
        \label{table:quant_comparison}
    \vspace{-5mm}
\end{table*}

\section{Dataset}
\label{sec:4}
We propose a novel dataset that is accurately annotated to train a reliable and effective model to predict violin performance motion from audio. The dataset includes audio recordings of the violin, comprehensive body motion data, and detailed information on bowing and fingering techniques. To prepare this dataset, we recruited three professional violinists. 
These musicians annotated bow up/down and finger numbers on the musical score and performed the pieces in a motion capture room, using predefined bowing and fingering techniques.

To acquire motion data, we employed a motion capture system that utilizes 16 synchronized cameras operating at 120 fps. The performer wore a suit with 57 optical markers provided by Vicon \cite{merriaux2017study}
, and an additional 20 markers were attached to the fingers. This resulted in the collection of the 3D positions of each marker over time, providing a detailed representation of the violinist's finger and body motions during the performance. 
Additionally, we simultaneously recorded audio and MIDI data using a violin that can output both audio and MIDI signals during motion capture. The audio signals were recorded at a sampling rate of 44.1 kHz, with care taken to eliminate any room noise. The MIDI data was recorded for the post-processing stage. 

\begin{figure}[tbp]
\vspace{-1.5mm}
\begin{minipage}[t]{0.48\linewidth}
    \centering
    \includegraphics[width=0.999\linewidth]{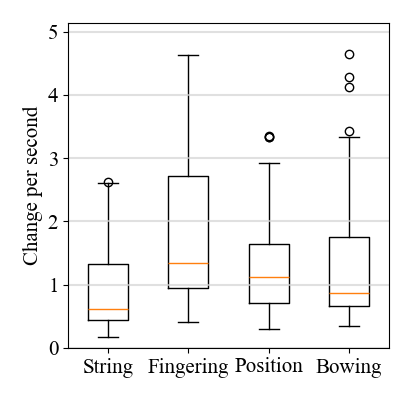}
    \vspace{-9mm}
    \caption{Statistics of the proposed dataset.}
    \label{fig:motion_analysis}
\end{minipage}%
    \hspace{0.04\linewidth}%
\begin{minipage}[t]{0.48\linewidth}
    \centering
    \includegraphics[width=0.999\linewidth]{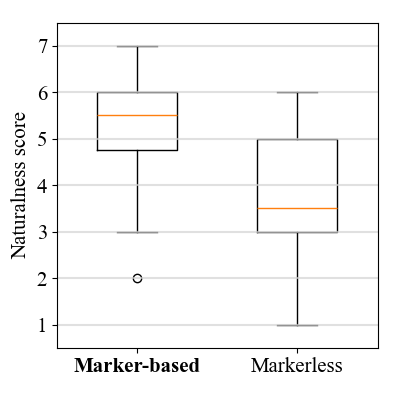}
    \vspace{-9mm}
    \caption{Subjetive evaluation of motion naturalness between marker- and markerless-based datasets.}
    \label{fig:sub_naturalness}
\end{minipage} 
\vspace{-6mm}
\end{figure}

Along with the motion and audio data, we obtained well-annotated information on fingering and bowing techniques for each performance. The performers entered this information into the music score data (in MusicXML format) before the recording, indicating the bow up/down and finger number for each note. 
In total, we collected 182.5 minutes of performance data, 61 pieces, from three violinists. 
We analyze the captured motions and diversity of the proposed dataset and present the statistics of hand motions in Fig.~\ref{fig:motion_analysis}. 
An extended analysis with tempo, pitch, and key distribution is available in the supplementary material.

Table~\ref{table:dataset_comparison} shows that the proposed dataset includes the most pieces and tracked joints among marker-/sensor-based datasets. In comparison to markerless datasets~\cite{jin2024audio, Shlizerman:18, Kao:20} with estimated joints based on vision, an extensive number of markers ensures our dataset of high reliability and accuracy. Additionally, using non-intrusive markers on fingertips provides valuable ground truth motion data for performers' hands, which is not available in existing work~\cite{volpe2017multimodal}. While marker-based methods are generally more accurate, they are often criticized for potentially affecting the naturalness of motion during capture. To address this, we conducted a subjective evaluation with 10 violinists, comparing the naturalness of motion between our marker-based dataset and the state-of-the-art markerless dataset~\cite{jin2024audio}. As shown in Fig.~\ref{fig:sub_naturalness}, 
the marker-based approach scored higher with a significance of $p<.001$.

To prepare the data for training our model, we performed post-processing on the recorded motion, audio, and bowing/fingering information. Detailed descriptions of the post-processing are included in the supplementary material. The final outputs include bow direction $L^{\mathrm{bow}}$, played string $L^{\mathrm{str}}$, finger number $L^{\mathrm{fing}}$, and position sequence $L^{\mathrm{pos}}$.



\section{Evaluation}

To verify the effectiveness of our proposed approach to accurately estimate violin performance motion from audio, 
we conducted comprehensive quantitative and subjective evaluations. We conducted experiments on a dataset consisting of 3.1 hours of violin performance data, collected as described in Section~\ref{sec:4}.
The data was split into a training set (54 pieces; 125 minutes), a validation set (8 pieces; 28.3 minutes), and a test set (8 pieces; 29.2 minutes). 
We normalized the audio signals so that the temporal mean of all channels is zero and the variance is one in the training set. Similarly, we normalized the motion data to ensure the temporal mean of all joints is zero, and the variance is one.

We employed three methods as the baseline
    that generates musical instrument performance motion 
    from audio signals similar to our method.
The baselines include
    the LSTM-based method ~\cite{Shlizerman:18},
    a two-branch network-based method ~\cite{Kao:20},
    and a GAN-based method~\cite{Chen:21}.
While the first two methods output joint positions similar to ours,  
    the last one outputs joint rotations.
Therefore, we calculated the joint positions based on forward kinematics from the output for a fair quantitative evaluation.


\subsection{Implementation details}
For each of the three convolutional layers in our bowing/fingering module, 
    the numbers of output channels were 32, 64, and 128,
    and the kernel sizes were $1 \times 1$, $3\times 3$, and $3\times 3$.
    The max-pooling kernel sizes were $4\times1$, $2\times1$, and $2\times 1$, respectively. 
The dimension of the BiLSTM layers was 512, and the dropout rate was set to 0.3. 
The output dimensions of the two FC layers were 64 
    and the number of output classes, respectively.  
The batch size was set to 8 for training the bowing/fingering module.

In the motion generation module,
    the output dimensions of the first FC layers
    to embed the bowing/fingering information and the audio features
    were 16 and 128, respectively.
The number of BiLSTM layers for generating left hand, left arm, right hand \& arm, and others
 were 2, 2, 2, and 3, 
 with dimensions of 256, 256, 512, and 128.
The dropout rate was also set to 0.3.
The batch size was set to 32 to train the motion generation module,
    and $\lambda$ in Eq.~\ref{eq: ce} was empirically set to 0.3 after experimenting.
The Adam solver ~\cite{Kingma:14adam}
    was used to optimize all model parameters, with a learning rate of 0.001.
All models were implemented using PyTorch~\cite{Pytorch}.

\subsection{Quantitative evaluation}

To evaluate the generated motions, we used three metrics: L1 loss, dynamic time warping (DTW) distance, and motion jerk. 
The L1 loss was calculated based on the difference between the joint positions of prediction and ground truth. The DTW distance was used to measure the similarity between the two trajectories of the predicted and ground-truth joint positions. We specifically evaluated each measure for all joints, the right arm (elbow and wrist), the left arm, and the fingers of the left hand, to assess different body parts that dominate the bowing and fingering. We further used jerk to quantify the smoothness of the generated motion data~\cite{Roren2022} by calculating the rate of change in the acceleration over all joints over time.

\begin{table}[t]
\vspace{1.5mm}
    \centering
    \resizebox{0.9\columnwidth}{!}{%
    \begin{tabular}{lcccc}
    \toprule
     Method & L1 & L1RA & L1LA & L1LF\\
     \midrule
      Shlizerman et al.~\cite{Shlizerman:18} & 18.08 & \bf{0.50} & 0.41 & 1.22 \\ 
Kao and Su~\cite{Kao:20} & 37.50 & 1.93 & 0.44 & 1.30\\ 
Chen et al.~\cite{Chen:21} & 20.02 & 0.74 & 0.34 & 1.17\\ 
\bf{Ours} & \bf{17.53} & 0.51 & \bf{0.38} & \bf{1.15}\\
\bottomrule
    \end{tabular}}
    \vspace{-2mm}
        \caption{Cross-violinist validation results. Mean scores are used to mitigate the variations between pieces.}
        \label{table:cross_violinist}
\vspace{-5mm}
\end{table}

Table \ref{table:quant_comparison} shows the quantitative results.
\{L1, DTW\} indicate the measures for all joints,
    \{L1RA, DTWRA\} indicate the measure for the right arm,
    \{L1LA, DTWLA\} indicate the measure for the left arm,
    and \{L1LF, DTWLF\} indicate the measure for the fingers of the left hand.
Each score represents the average of the entire test set. To ensure a fair comparison, we made only minimal adjustments to the input and output dimensions of the previous models.
As shown in the table, our method outperformed the baseline methods across almost all scores,
 except for a slight shortfall in two metrics compared to one method~\cite{Chen:21}.
These results indicate that our method is more effective in generating accurate and smoother body motions than all baseline methods. 
To verify network generalizability, we further conducted a cross-violinist evaluation by removing each performer from the training data and evaluating accuracy with the absent performer's data. Table~\ref{table:cross_violinist} shows a good generalization of our method even in the case of unseen styles.

\subsection{Subjective evaluation}
\label{subsec:sub_eval}

\begin{figure*}[t]
\includegraphics[width=0.9\linewidth]{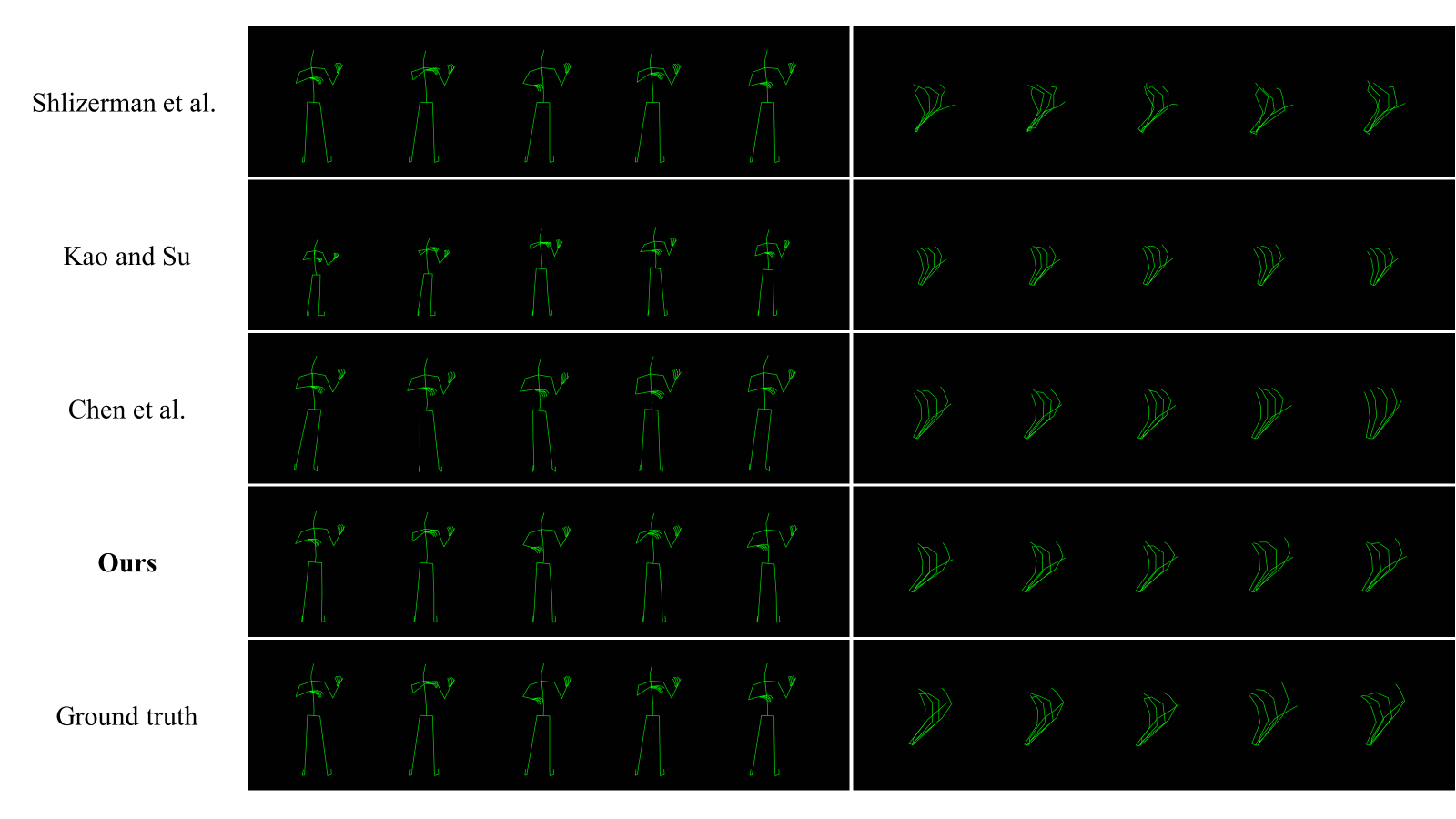}
  \vspace{-4mm}
  \caption{Samples of generated results and ground truth. More results are available in the supplementary material.}
  \vspace{-4mm}
\label{fig:result1}
\end{figure*}

We conducted a subjective evaluation to assess the naturalness of the motion generated by our method (see Fig.~\ref{fig:result1}). 
We randomly selected 20-second segments from each of the eight pieces in the test data and created video clips with the generated results. 
An online form was created, presenting participants with the generated body motions without rigged avatars to ensure a clear comparison between methods.
A total of 46 participants took part in our subjective evaluation, 
 including 4 professional violinists, 16 amateurs, and the remainder with no prior experience playing the violin.
After watching each video clip, participants rated the naturalness of the motion on a seven-point Likert scale (ranging from 1: very unnatural to 7: very natural).

\begin{figure}[t]
    \centering
    \includegraphics[width=0.8\linewidth]{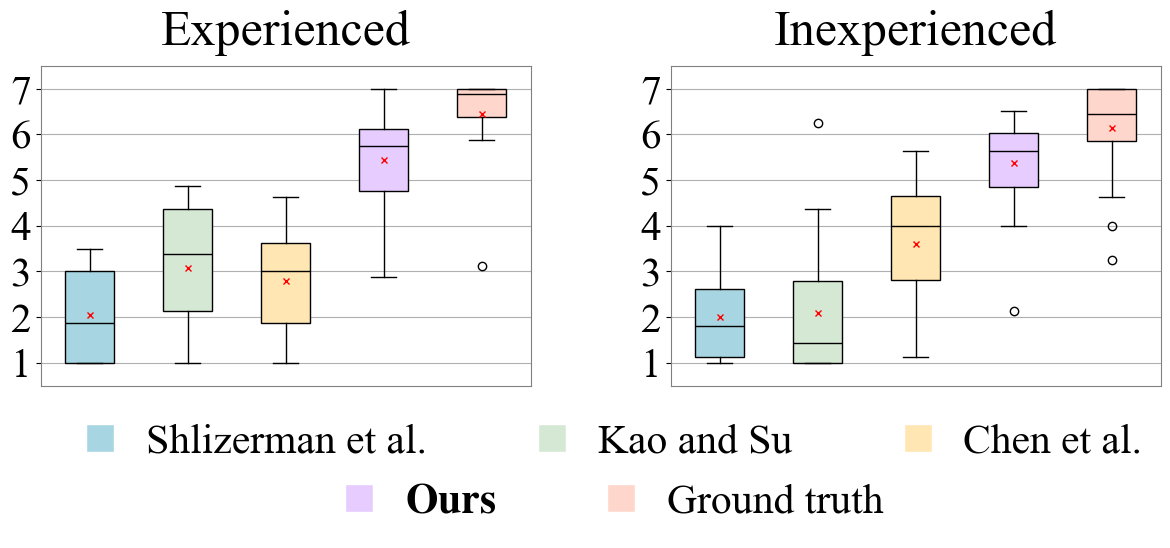}
    \vspace{-3mm}
    \caption{Subjective evaluation results in comparison.}
    \label{fig:sub_comparison}
\vspace{-6mm}
\end{figure}

Figure~\ref{fig:sub_comparison} presents the box plots and the average rating scores, divided between participants with and without violin experience. 
Table~\ref{table:sub_comparison} shows the result of a Wilcoxon signed-rank test. 
It demonstrates that our method significantly outperformed all three existing methods in terms of naturalness for both experienced and inexperienced participants,
although experienced participants tend to show a larger disparity between the scores given to the ground truth and the generated results.
For a more detailed comparison, please refer to the supplementary material.



\subsection{Ablation study}



To evaluate the effect of each feature within the bowing/fingering information, we trained four different motion generation modules, each excluding one of the four features: bow direction, played string, finger number, and position. We then fed the estimated bowing/fingering information from the audio, with each feature excluded, into the motion generation module and analyzed the generated results in the ablation study. Moreover, to assess the elaborate motion generation module with separate body semantics, we compared it against a configuration using only a single branch of BiLSTM. 
This single branch BiLSTM comprises two BiLSTM layers with 512 dimensions,
 where the number of layers and dimensions were experimentally determined for optimal performance.
Finally, to evaluate our loss design, an ablation study of $L_\mathrm{dis}$ is shown in Table~\ref{table:quant_ablation}.

 \begin{table}[t]
 \vspace{1mm}
    \centering
    \resizebox{0.9\columnwidth}{!}{%
    \begin{tabular}{lcc}
    \toprule
    \multicolumn{1}{l}{\multirow{2}{*}{Tested method}} & \multicolumn{2}{c}{$p$-value} \\
    \cmidrule(lr){2-3}
       & Experienced & Inexperienced \\
     \midrule
      Shlizerman et al.~\cite{Shlizerman:18} & $0.002$*  & $2.861 \times 10^{-5}$*    \\
      Kao and Su~\cite{Kao:20}               & $0.002$*  & $5.202 \times 10^{-5}$*	 \\
      Chen et al.~\cite{Chen:21}	           & $0.002$*  & $5.223 \times 10^{-5}$*    \\
      Ground truth	                       & $0.002$*  & $6.373 \times 10^{-5}$*    \\
\bottomrule
    \end{tabular}}
        \vspace{-1mm}
        \caption{Wilcoxon signed-rank test results for the scores of all baseline methods and the ground truth
        compared with the proposed method.
        ``*'' indicates the 0.0125 significance level, corrected from 0.05 using the Bonferroni method 
        to prevent an increase in Type I error during multiple comparisons.}
        \label{table:sub_comparison}
    \vspace{-4mm}
\end{table}


\begin{table*}[t]
    \centering
    \resizebox{1.8\columnwidth}{!}{%
    \begin{tabular}{lccccccccc}
    \toprule
     Condition & L1 & L1RA & L1LA & L1LF & DTW & DTWRA & DTWLA & DTWLF & Jerk\\
     \midrule
Without $L^{\mathrm{bow}}$ &+0.8004	&+0.0194 &+0.0252 &+0.0414 &+0.9164	&+0.0337 &+0.0184 &+0.0412  & +6.65\\
Without $L^{\mathrm{str}}$ &+1.6392	&+0.0412 &+0.0499 &+0.0968 &+2.1172 &+0.0527 &+0.0595 &+0.1460 & +8.31\\
Without $L^{\mathrm{fing}}$ &+0.8886 &+0.0401 &+0.0117 &+0.0030	&+1.4702 &+0.0647 &+0.0181 &+0.0169 & +40.83\\
Without $L^{\mathrm{pos}}$ &+0.1146 &+0.0046 &-0.0018 &+0.0006 &-0.0618 &+0.0009 &-0.0065 &-0.0142& +8.25\\
\midrule
Single branch &+0.4437 &+0.0221 &+0.0019 &-0.0090 &+0.3879 &+0.0150 &-0.0033 &-0.0096 & +538.05\\
\midrule
Without $L_{\mathrm{dis}}$ &+1.8348 &+0.0538 &+0.0466 &+0.0826 &+2.6668 &+0.1002 &+0.0640 &+0.1060 &+81.63\\
\bottomrule
    \end{tabular}}
        \vspace{-2mm}
        \caption{Quantitative results from the ablation study. A larger positive value indicates a greater contribution of each feature to the score.}
        \vspace{-4mm}
    \label{table:quant_ablation}
\end{table*}

\subsubsection{Quantitative evaluation results}

Regarding the ablation study, the top four rows of Table~\ref{table:quant_ablation} show the difference between the results when each feature was excluded and when all features were used. A larger positive value in this table indicates a greater contribution of the removed feature to each score. 
The results reveal that most values were positive,
 with the bow direction $L^{\mathrm{bow}}$ having a moderate impact even on the DTW distance for the right arm.
The played string $L^{\mathrm{str}}$ contributed the most to L1 and DTW scores, especially the DTW distance for the left-hand fingers.
The finger number $L^{\mathrm{fing}}$ notably improved the jerk score.
Unlike these three features,
 the contributions of position $L^{\mathrm{pos}}$ were relatively small. 
The bottom row shows the difference between the results of the single- and multi-branch BiLSTM,
 indicating that our semantics-aware module significantly improved the jerk score.
In summary, the results show that the bowing and fingering information contributed to the body parts related to each feature, 
 and a body semantics-aware module outperformed a single-branch BiLSTM with improved naturalness.

As there was no established benchmark for estimating bowing and fingering from audio, we report the following estimation accuracies: bow direction, played string, finger number, and position
    for 0.724, 0.942, 0.664, and 0.759.


\begin{figure}[t]
    \centering
    \includegraphics[width=0.8\linewidth]{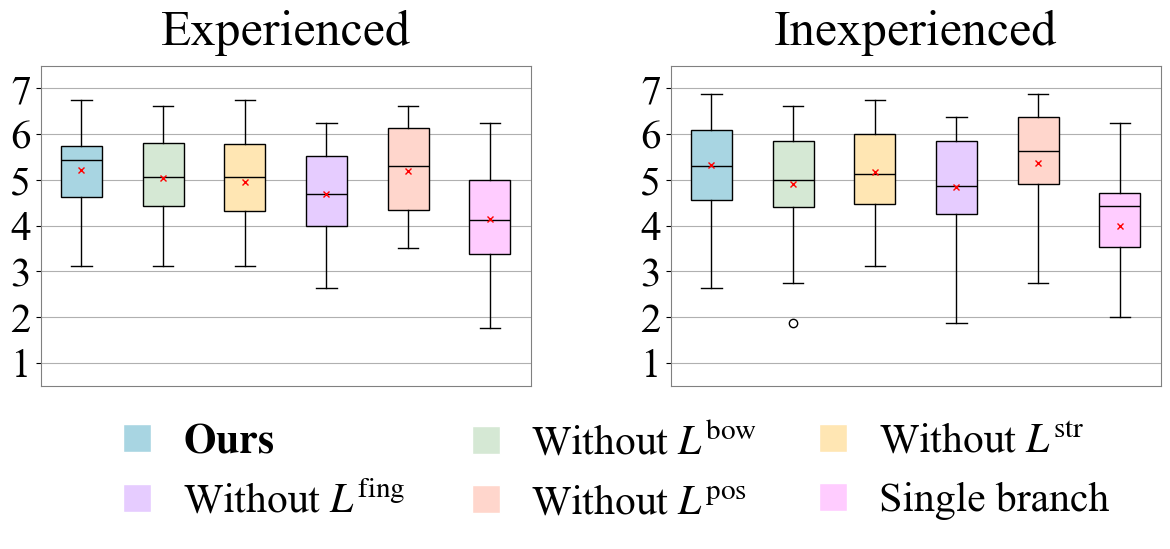}
    \vspace{-3mm}
    \caption{Subjective evaluation results from the ablation study.}
    \label{fig:sub_ablation}
    \vspace{-5mm}
\end{figure}

\subsubsection{Subjective evaluation results}
Figure~\ref{fig:sub_ablation} and Table \ref{table:sub_ablation} present the results from a subjective evaluation identical to Section \ref{subsec:sub_eval}.
For experienced participants, 
 the proposed method demonstrated improved naturalness in motion compared to the models 
 lacking the played string feature $L^{\mathrm{str}}$ and the finger number feature $L^{\mathrm{fing}}$.
For inexperienced participants,
 naturalness improved compared to the models
 without the bow direction feature $L^{\mathrm{bow}}$ and the finger number feature $L^{\mathrm{fing}}$.
These results suggest that experienced participants focused on the details of the left hand, 
 while inexperienced participants rather focused more on the movements of the right hand.
This may be because violin players often monitor their left hand while playing 
 to ensure their fingers are pressing the correct strings, 
 making them unconsciously more attentive to left-hand details. 
In contrast, inexperienced participants notice more prominent visual features, 
 with the right hand's movement being the most visually apparent.
Additionally, both groups noticed the difference in naturalness 
 between the proposed model and the single-branch model, 
 indicating that our body semantics-aware architecture significantly contributed to a higher level of realism.

\begin{table}[t]
    \vspace{1mm}
    \centering
    \resizebox{0.9\columnwidth}{!}{%
    \begin{tabular}{lcc}
    \toprule
     \multicolumn{1}{l}{\multirow{2}{*}{Condition}} & \multicolumn{2}{c}{$p$-value}  \\
    \cmidrule(lr){2-3}
      & Experienced & Inexperienced \\
     \midrule
Without $L^{\mathrm{bow}}$  & $0.078$                   & $3.409 \times 10^{-4}$**  \\
Without $L^{\mathrm{str}}$  & $0.002$**                 & $0.106$                   \\
Without $L^{\mathrm{fing}}$ & $0.004$**                 & $3.357 \times 10^{-4}$**  \\
Without $L^{\mathrm{pos}}$  & $0.886$                   & $0.808$                   \\
\midrule
Single branch           & $1.406 \times 10^{-4}$**  & $5.352 \times 10^{-5}$**  \\
\bottomrule
    \end{tabular}}
    \vspace{-2mm}
    \caption{Wilcoxon signed-rank test results for the models of the ablation study.
        ``**'' indicates the 0.01 significance level, corrected from 0.05 using the Bonferroni method.}
    \label{table:sub_ablation}
    \vspace{-6mm}
\end{table}

\section{Conclusion and Discussion}

In this paper, 
 we presented a novel multi-stage approach for generating motion from audio in violin performance.
Diverging from previous methods that focused solely on predicting joint positions or rotations from audio, 
 our approach integrated a bowing/fingering module and a motion generation module. 
This integration allowed us to reflect the violinist's bowing and fingering intentions in the resulting motion.
Experimental results demonstrated that our model achieved greater precision and naturalness compared to previous methods.
Additionally, we created a new violin performance dataset 
 comprising audio signals, accurately captured corresponding motion data, 
 and well-annotated bowing/fingering information. 
Upon publication, 
 we intend to release this multimodal dataset to the research community for further exploration.


The current approach is limited in generating artist-specific motion, as individual interpretations heavily influence bowing and fingering choices.
Future work will explore capturing performers' unique styles for greater expressiveness and customization.
Moreover, we plan to extend our multi-stage approach to other string instruments, such as cellos and guitars, which also require nuanced global and fine-grained motions.



\clearpage

{\small
\bibliographystyle{ieee_fullname}
\bibliography{refs}
}


\end{document}